\documentclass[aps,prl,twocolumn]{revtex4}
\usepackage{epsfig}
\begin{document}

\title{The large-scale angular correlations in CMB temperature maps}

\author{Armando Bernui\,$^{\ast}$}
\affiliation{Instituto Nacional de Pesquisas Espaciais \\
             Divis\~{a}o de Astrof\'{\i}sica  \\
Av. dos Astronautas 1758, 12227-010 -- S\~ao Jos\'e dos Campos, SP, Brazil}


\begin{abstract}
Observations show that the Cosmic Microwave Background (CMB) contains 
tiny variations at the $10^{-5}$ level around its black-body equilibrium 
temperature. 
The detection of these temperature f\/luctuations provides to modern 
Cosmology evidence for the existence of primordial density perturbations 
that seeded all the structures presently observed. 
The vast majority of the cosmological information is contained in the 
2-point temperature function, which measures the angular correlation of 
these temperature f\/luctuations distributed on the celestial sphere. 
Here we study such angular correlations using a recently introduced 
statistic-geometrical method.
Moreover, we use Monte Carlo simulated CMB temperature maps to show the 
equivalence of this method with the 2-point temperature function (best 
known as the 2-Point Angular Correlation Function). 
We also investigate here the robustness of this new method under 
possible divisions of the original catalog-data in sub-catalogs. 
Finally, we show some applications of this new method to simple cases.
\end{abstract}

\pacs{98.80.-k, 98.65.Dx, 98.70.Vc}
\keywords{cosmology: large-scale structure of the universe --- CMBR, anisotropies}

\maketitle

%
\section{Introduction}

According to modern Cosmology, in the primordial Universe ionized 
barionic matter and photons were tightly coupled until the recombination 
epoch when the hydrogen and helium nuclei hold their electrons to form
neutral atoms, leaving photons to freely stream toward us. 
These photons, nowadays observed as a faint residual relic radiation  
termed the Cosmic Microwave Background Radiation (CMB), 
have a well described black-body Planckian spectrum as the FIRAS 
instrument on the COBE satellite showed in 1992 (see e.g.~\cite{COBE} 
and references therein).

Besides this thermal equilibrium feature, the CMB preserved evidences 
of small variations in temperature (or intensity) from one part of the 
microwave sky to another, for this called CMB temperature f\/luctuations 
(also known as CMB {\em anisotropies} in the literature), 
which are interpreted as being originated during their interaction 
with the primordial matter-density f\/luctuations. 
Since at that time the density f\/luctuations were correlated at 
specific angular scales characteristic of the physical processes 
involved (e.g.,~\cite{Hu,Hu2}), such correlations were imprinted in 
the angular distribution of the the CMB temperature f\/luctuations. 
This fact turns of fundamental importance the study of the CMB angular 
correlations, in order to understand the evolution of the 
matter-density f\/luctuations to form the structures --like galaxies 
and galaxy clusters-- observed today. 

In the last decades, especially after the COBE mission, observational 
cosmology evolved astonishingly fast, in such a way that there has been 
an increasing improvement in the quality of measuring the CMB 
temperature f\/luctuations. 
While with COBE data the CMB studies were restricted to large 
angular-scales, with the highly precise and excellent angular 
resolution of WMAP data, analysis at all angular-scales are now 
possible.

In what follows we present the basics of a recently proposed 
{\em method} suitable for the study of the angular correlations in 
the sky distribution of cosmic objects. 
Our main interest is focussed in CMB data like those released by 
the WMAP satellite~\cite{WMAP}, for this we shall apply the method 
to CMB temperature f\/luctuations maps (brief\/ly {\bf CMB maps}), 
and also investigate the robustness of the method under possible 
partitions of the original data in sub-catalogs. 
Finally, we shall show the equivalence of our method with the 2-Point 
Angular Correlation Function (2-PACF), and this will be done both in 
a theoretical way using the definitions of these approaches as well 
as in a computational way using Monte Carlo simulated CMB maps. 

\vspace{-0.4cm}
%
\section{The PASH method}

Recently, Bernui \& Villela~\cite{BV, BVF} introduced a new 
method to scrutiny the angular correlations in the distribution of 
cosmic objects in the celestial sphere. 

\noindent
Given a catalog with $n$ cosmic objects (sharing common physical 
properties), the method consists in first calculate the $n(n-1)/2$ 
angular distances of all pairs-of-objects and then construct the 
normalized histogram of the number of pairs with a given angular 
separation {\it versus} their angular separation. 
The Expected-PASH (EPASH) is the histogram obtained for the 
ideal distribution case. 
The dif\/ference between the PASH, caculated from the observational 
data, and the EPASH can reveal significant departures of the former 
histogram with respect to the later, if the signal-to-noise ratio 
is sufficiently large. 
This geometric-statistical tool is called the Pair Angular 
Separation Histogram (PASH) method. 

\vspace{0.2cm}
Catalogs with a large number of objects, as high-resolution CMB 
maps, can be suitable divided in a set of comparable sub-catalogs, 
provided that they contain a similar number of objects sharing 
analogous physical properties. 
After the partition of the original catalog in, say $K$, sub-catalogs 
one performs a PASH with each one of these sub-catalogs and average 
them to obtain the Mean-PASH (MPASH).
As we shall see below, the MPASH function is independent of the 
number of sub-maps $K$ in which the original map is divided, 
provided that the minimum number is  $K_{\rm{min}}=2$: 
one sub-map for the positive CMB and the other for the negative 
ones.
Finally we perform, and plot, the dif\/ference between the MPASH and 
the EPASH. 

\vspace{0.2cm}
The method has the advantage that it does not depend on cosmological 
parameters or data other than the angular position in the celestial 
sphere of the cosmic objects listed in the catalog to be analised. 
The method is also applicable to incomplete sky maps, including 
disconnected regions, on condition that the EPASH was obtained under 
similar characteristics (i.e. the same number of objects per catalog, 
and where objects are located in the same patch of the celestial 
sphere).
Notice that the WMAP CMB maps do not include the monopole and dipole 
contributions, therefore for such data the EPASH has to obtained 
taking into account these features.

\vspace{0.2cm}
We firstly describes the basics of the method by showing how to 
obtain a PASH given a catalog of data.
A catalog $C$ is a list of $n$ cosmic objects containing suitable 
information about their angular coordinates on the celestial sphere
and some physical properties; 
here we assume that the objects of a given catalog have common 
physical properties. 

\noindent
We divide the interval $(0,180^{\circ}]$ in $N_{bins}$ bins 
of equal length $\delta \gamma \,=\, 180^{\circ} / N_{bins}$, where each 
sub-interval has the form 
$J_i = (\gamma_{i}^{} - \frac{\delta \gamma}{2} \,,$
$\gamma_{i}^{} +\frac{\delta \gamma}{2}] \,,\, 
i\!=\!1,2, \dots ,N_{bins} \, ,$ with center in
$\gamma_{i}^{} = \,(i - \frac{1}{2}) \,\, \delta \gamma \,$.
Next, we calculate the $N \equiv n(n-1)/2$ angular distances 
corresponding to the distances between all pairs-of-objects. 
We denote by $\varphi(\gamma)$ the number of pairs of objects
in $C$ separated by a distance $\gamma \in (0,180^{\circ}]$.
Then, 
\begin{equation} 
\Phi(\gamma_{i}^{}) = \frac{1}{N\,\delta \gamma} \,\,
\sum_{\gamma \in J_i} \varphi(\gamma)  \, ,
\end{equation}
is the normalized counting of the number of pairs of objects, 
i.e. $\sum_{i=1}^{N_{bins}} \Phi(\gamma_{i}^{})\, \delta \gamma = 1 \,$,
separated by an angle $\gamma_{i}^{}$, that lies in the 
sub-interval $J_i$. 

\noindent  
Let $p, \,q$ any two objects $p, \,q \in C$, with angular coordinates 
$(\theta_p, \phi_p), \, (\theta_q, \phi_q)$, respectively,  
where $\theta_p,\, \theta_q \in [0,180^{\circ}]$ and  
$\phi_p,\, \phi_q \in [0,360^{\circ}]$.
The unit vectors $\vec{n}_p, \,\vec{n}_q$ (i.e., $|\vec{n}_p| = |\vec{n}_q| = 1$) 
denote the position of the two objects $p, \,q$ on the celestial 
sphere. 
We denote by $\rho(\vec{n}_p)$ the probability densities of the 
objects distributed on the celestial sphere ${\cal S}^2$ and listed 
in $C$. 
Then, the probability density that two objects $p, q \in C$
be separated an angular distance $\gamma \in [0,180^{\circ}]$ is 
defined by 
\begin{eqnarray} \label{def-PASH}
{\cal P}(\gamma) \equiv \int \!\! \int \, 
d\Omega_{p}\, d\Omega_{q}\, \rho(\vec{n}_p)\, \rho(\vec{n}_q) \,
\delta(\, \gamma - \mbox{\rm \^{d}}(\vec{n}_p,\vec{n}_q) \,)  \, , 
\end{eqnarray}  
where $\delta$ is the Dirac-delta, and  
$\mbox{\rm \^{d}}(\vec{n}_p,\vec{n}_q) = \arccos \, [ \cos\theta_p \cos \theta_q 
+ \sin \theta_p \, \sin \theta_q \cos(\phi_p - \phi_q) ]$.

\noindent
Hence, the PASH is simply defined by 
\begin{equation} 
\Phi(\gamma_{i}^{}) \equiv \frac{1}{\delta \gamma} \,
\int_{J_i} {\cal P}(\gamma) \, d\gamma \, .
\end{equation}
However, if the interval $J_i$ is small enough (as shall be
considered here), a suitable approximation for the PASH is
\begin{equation} 
\Phi(\gamma_{i}^{}) \simeq {\cal P}(\gamma_{i}^{})\, .
\end{equation}

\noindent
Notice that $\rho(\vec{n}_p) d\Omega_{p}$ is the 
probability of the object $p$ to be in the sky 
patch $d\Omega_{p} \in S^2$.
Of course, the probability of finding $p$ in the whole 
$S^2$ should be equal to 1,
\begin{eqnarray}
\int_{S^2}\, \rho(\vec{n}_p) \, d\Omega_p = 1\, . \nonumber
\end{eqnarray}
Thus, a purely isotropic and normalized density distribution
of objects observed in the full-sky $S^2$, reads
$\rho(\vec{n}_p) = \frac{1}{4 \pi}$. 
With this information, a direct calculation in eq.(\ref{def-PASH})
gives,
\begin{eqnarray} \label{EPASH}
{{\cal P}_{_{\mbox{\footnotesize full-sky}}}}
^{\hspace{-0.95cm} 
\mbox{\footnotesize Expected}}\,(\gamma) = \frac{1}{2} \sin \, \gamma \, ,
\end{eqnarray}
is the EPASH for the case of isotropically distributed objects 
on the full-sky $S^2$.

\noindent
Notice that a PASH ${\cal P}(\gamma)$  satisfies the normalisation 
property: $\int_{0}^{180^{\circ}} {\cal P}(\gamma) \, d\gamma \,=\, 1$, 
which let us to perform the mean of an arbitrary number of histograms, 
for sub-catalogs containing comparable number of objects. 
This is a very useful condition since produces a sort of 
{\em normalisation} of our method in the sense that {\em area} of 
the function MPASH-minus-EPASH is zero: 
\begin{eqnarray} \label{zero-area}
\int_{0}^{180^{\circ}} 
(\, {\cal P}^{\mbox{\footnotesize MPASH}}(\gamma)
   -{\cal P}^{\mbox{\footnotesize EPASH}}(\gamma) \,) 
\, d\gamma \,=\, 0.
\end{eqnarray}

%
\section{The 2-Point Angular Correlation Function}

\noindent
The Two-Point Angular Correlation Function (2-PACF) is defined by 
\begin{eqnarray} \label{2PACF}
C(\gamma) \equiv \langle \,T(\vec{n}_p) \, T(\vec{n}_q)\, \rangle \,, 
\end{eqnarray}
where $\vec{n}_p, \vec{n}_q$ are such that 
$\cos\gamma = \vec{n}_p \cdot \vec{n}_q$, 
and $T(\vec{n}_p), \, T(\vec{n}_q)$ are the temperature f\/luctuations 
of the pixels $p, \,q$ of a given CMB map, respectively. 
Notice that, in some literature (see, e.g.~\cite{Bersanelli}) this 
definition appears with $T(\vec{n}_p)/T_0$ instead of $T(\vec{n}_p)$, 
where $T_0$ is the black-body equilibrium temperature of the CMB. 

\noindent
Since $\vec{n}_i, \,i=p,q$ is a vector with angular coordinates
$(\theta_i,\phi_i), \,i=p,q$, one can expand $T(\vec{n}_i)$ 
in the spherical harmonics, so 
\begin{eqnarray} \label{Texp}
T(\vec{n}_i) 
= \sum_{\ell\, m} a_{\ell\, m} Y_{\ell\, m}(\theta_i,\phi_i) \, ,
\end{eqnarray}
where $a_{\ell\, m}$ are the multipole moments, with zero mean 
$\langle a_{\ell\, m} \rangle = 0$ for Gaussian temperature 
f\/luctuations. 
In a statistically isotropic Universe the variance of the multipole 
moments is independent of $m$, which means that we can define
\begin{eqnarray} \label{Cl}
{\cal C}_{\ell} \equiv 
\frac{1}{2\ell+1} \sum_{m={\mbox{\small -}}\ell}^{\ell} \, |a_{\ell\, m}|^2
\, ,
\end{eqnarray}
and the set of ${\cal C}_{\ell}$ is termed the angular power spectrum 
of the CMB map.

Using eqs. (\ref{Texp}) and (\ref{Cl}), and well-known properties of 
the spherical harmonics in the definition (\ref{2PACF}), the 2-PACF 
for a statistically isotropic Universe can be written as
\begin{eqnarray} \label{C-gama}
C(\gamma) =
\frac{1}{4\pi} \sum_{\ell}(2\ell+1) {\cal C}_{\ell} P_{\ell}(\cos\gamma)\,,
\end{eqnarray}
where $P_{\ell}$ are the Legendre polynomial of oreder $\ell$.
We define the function 
\begin{eqnarray}
C^{norm}(\gamma)\equiv \frac{1}{2}\sin\gamma \,C(\gamma) \, ,
\end{eqnarray}
and call it the {\em normalized} 2-PACF because the area between 
the curve and the horizontal axis is zero, as can be verified from
the bottom-plot appearing in Figure 1.
Moreover, assuming a Harrison-Zeldovich (HZ) scale-invariant spectrum 
and for large angular scales (i.e., $\ell \lesssim 100$) we have 
${\cal C}_{\ell} = \mbox{Q}/(\ell(\ell+1))$, where $\mbox{Q}$ is the 
quadrupole normalisation constant~\cite{Bersanelli,Padmanabhan}.
In Figure 1 we plotted both functions, $C(\gamma)$ and $C^{norm}(\gamma)$,
for the case of a statistically isotropic Universe with HS spectrum.

\begin{figure} 
\includegraphics[width=9cm, height=12cm]{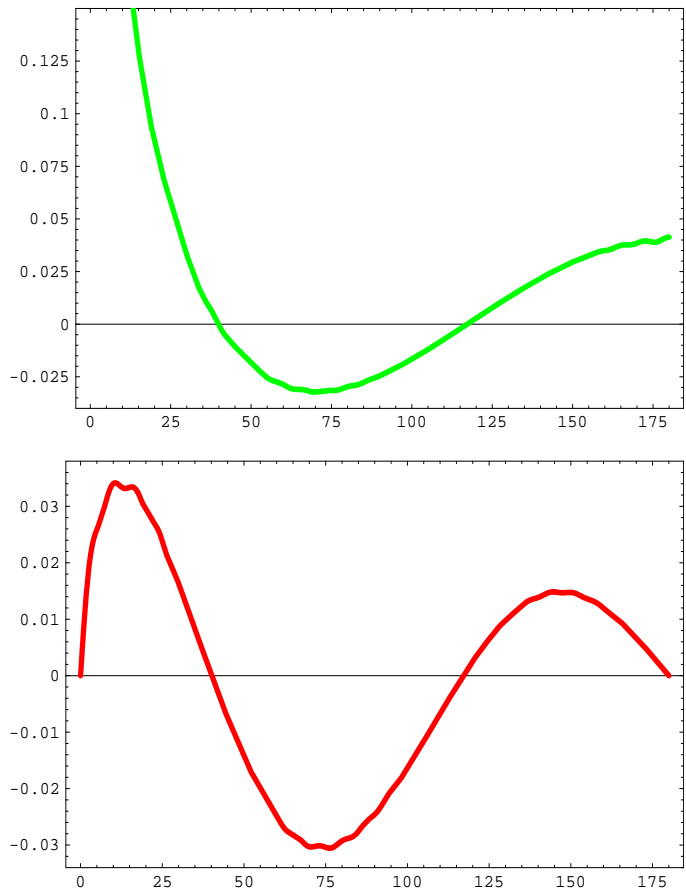}
\caption{ \label{figure1}
Here we consider a statistically isotropic Universe with HZ spectrum.
Moreover, motived in WMAP data where the monopole and dipole 
contributions are not included, here we consider that the sum index 
starts with $\ell=2$, i.e. with the quadrupole moment. 
{\bf Top}: We plotted the 2-PACF, given by $C(\gamma) 
= \frac{\mbox{Q}}{4\pi} \sum_{\ell=2} \frac{2\ell+1}{\ell(\ell+1)} 
P_{\ell}(\cos\gamma)$.
{\bf Bottom}: We plotted the {\em normalized} 2-PACF, which defined 
by $C^{norm}(\gamma) \equiv \frac{\sin\gamma}{2} \,C(\gamma)$;
this 2-PACF is called {\em normalized} because the area between the 
curve and the horizontal axis is zero.}
\end{figure} 

When we use the formula (\ref{C-gama}), instead of the definition
eq. (\ref{2PACF}),  a crucial point appears in the method of the 2-PACF. 
In fact, intrinsically to the method there is a large uncertainty 
in the evaluation of the ${\cal C}_{\ell}$'s, called the Cosmic 
Variance, and is given by 
$\delta {\cal C}_{\ell}/{\cal C}_{\ell} = \sqrt{2/(2\ell+1)}$.
This reveals no more than our ignorance in determining the real values 
of the low-order ${\cal C}_{\ell}$ (i.e. for $\ell=2,3,4,...$). 

%
\section{The 2-PACF {\footnotesize \it versus} the 
PASH-$\mbox{\rm\normalsize minus}$-EPASH function}

Here we shall prove the equivalence between the PASH method and 
the 2-PACF. 

The above definition of the 2-PACF, eq. (\ref{2PACF}), can also 
be written as (see~\cite{Padmanabhan}, pag. 195, eq. (5.28), 
where $\xi(\mbox{\bf x}) = \xi(\cos\gamma) \equiv C(\gamma)$)
\begin{eqnarray}
C(\gamma) \,+\, 1 && \nonumber
\end{eqnarray}

\vspace{-0.8cm}
\begin{eqnarray}
\!=\!A\!\!\int\!\! d\Omega_q \,\rho(\vec{n}_q)\!\int\!\! d\Omega_p \,\rho(\vec{n}_p)
\delta(\cos\theta_p\!-\!\cos(\gamma+\theta_q))\delta(\phi_p\!-\!\phi_q), \nonumber 
\end{eqnarray}
where $A$ is a constant, and $\delta$ is the Dirac-delta.
Multiplying both sides by $\frac{\sin\gamma}{2}$, which according 
to eq.~\ref{EPASH} is the EPASH for the full-sky catalog, and using 
properties of the Dirac-delta we obtain
\begin{eqnarray}
\frac{\sin\gamma}{2} \, C(\gamma) \,+ \, \frac{\sin\gamma}{2} && \nonumber 
\end{eqnarray}

\vspace{-0.8cm}
\begin{eqnarray} 
\,= \frac{A}{2} \! 
\int\! d\Omega_q\,\rho(\vec{n}_q)\!\int\! d\Omega_p \rho(\vec{n}_p)
\delta(\theta_p -\theta_q -\gamma)\delta(\phi_p -\phi_q), && \nonumber 
\end{eqnarray}
which means that when $\phi_p=\phi_q$, then $\gamma=\theta_p -\theta_q$;
in other words the only contribution to the integral comes from 
the equality $\gamma = \mbox{\rm \^{d}}(\vec{n}_p,\vec{n}_q)$. 
This implies that the right-hand-side of this equation is equal to 
the definition of the PASH, eq. (\ref{def-PASH}), except for the 
constant $A$. 
Thus, choosing suitably the value of $A$, we formally obtain the 
equivalence between both approaches
\begin{eqnarray}
\frac{\sin\gamma}{2} \,C(\gamma) \,=\, 
  {\cal P}^{\mbox{\footnotesize MPASH}}(\gamma)
- {\cal P}^{\mbox{\footnotesize EPASH}}(\gamma) \, . \nonumber
\end{eqnarray}

Moreover, we also performed here a computational verification of 
this equivalence. 
For this, we carry out the average of 1000 MPASH-minus-EPASH functions 
from an equal number of Monte Carlo CMB full-sky maps generated under 
the statistical isotropy and HZ spectrum assumptions. 
Our results, observed in Figure 2, show the excellent coincidence 
between both functions which is a computational verification 
of the equivalence between the MPASH-minus-EPASH function and the 
{\em normalized} 2-PACF.

Therefore, we conclude that the two approaches, the MPASH-minus-EPASH 
and the {\em normalized} 2-PACF 
($C^{norm}(\gamma)=\frac{\sin\gamma}{2}C(\gamma)$) are equivalent. 
However, it is worth to mention here that the {\em normalized} version 
of the 2-PACF seems to be more suitable than the original 2-PACF 
version, because it leds to scrutiny the angular correlations at 
all angular scales.  
As observed in the top-plot of Figure 1, the original 2-PACF is not 
well-defined for small angular scales $[0,\sim 20^{\circ}]$.  

\begin{figure} 
\includegraphics[width=9cm, height=7cm]{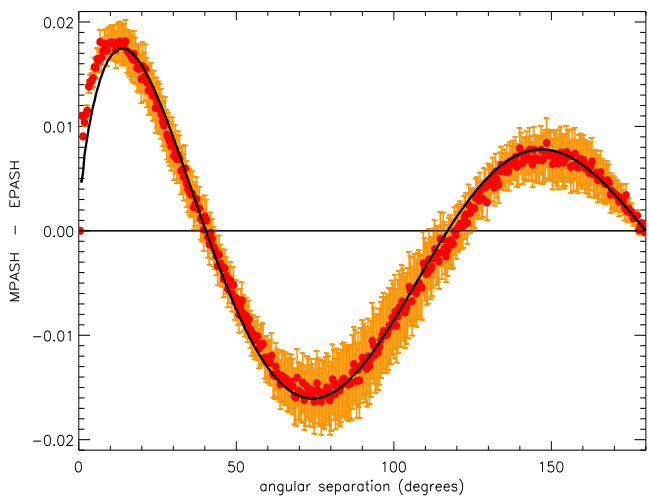}
\caption{ \label{figure2}
The red-points (including the standard deviation bars at $1 \sigma$)
correspond to average of 1000 MPASH-minus-EPASH functions from Monte 
Carlo CMB maps generated under the statistical isotropy assumption, 
and the black-line corresponds to the {\em expected} 2-PACF
$C^{norm}(\gamma)=\frac{\sin \gamma}{8\pi} 
\sum_{\ell=2} \frac{2\ell+1}{\ell(\ell+1)} P_{\ell}(\cos\gamma)$ 
for a statistically isotropic Universe. 
We realize that the excellent coincidence between both results is a 
validation of the equivalence between the MPASH-minus-EPASH function 
and the {\em normalized} 2-PACF.} 
\end{figure} 

%
\section{Robustness of the MPASH calculation}

Here we shall study the robustness in the obtention of the MPASH, 
in the case that it could be obtained considering different 
partitions (or ways to generate the sub-catalogs) of the original 
catalog (e.g. the temperature data in a CMB map). 
We shall deal with this problem by directly showing our results of 
calculating the MPASH in different partitions (different number of 
sub-catalogs $K$) of a given WMAP CMB temperature map.

A CMB full-sky map is the celestial sphere partitioned in a set 
of equal-area pixels, where to each pixel is assigned a temperature 
value; so the above mentioned physical property is the CMB temperature 
f\/luctuation value of each pixel. 
To obtain the angular correlations of a CMB map, we order the set 
of pixels according to increasing values of the pixels temperature. 
Next, we divide this ordered set of pixels in $K$ disjoint sub-sets 
(termed sub-catalogs), all of them with a comparable number of pixels. 
Afterward, we compute the MPASH averaging the $K$ PASHs calculated 
from each of the $K$ sub-catalogs. 
As we shall show in Figure 6, the MPASH function is independent of 
the number of sub-catalogs $K$ in which the original map is divided, 
provided that the minimum number is  $K_{\rm{min}}=2$: 
one sub-map for the positive CMB and the other for the negative 
ones.
However, it is clear that we can divide the original data in several 
ways, always following these criteria.

Here we consider the CMB map obtained using WMAP data, specifically 
the co-added Q+V+W map~\cite{WMAP2}, which is a weighted combination 
of the 8 high frequency dif\/ferencing assemblies (DA): 
Q1, Q2, V1, V2, W1, W2, W3, and W4 (see: 
http:$\backslash \backslash$lambda.gsfc.nasa.gov$\backslash$product
$\backslash$map$\backslash$current$\backslash$IMaps
$_{_{\mbox{--}}}$cleaned.cfm), that is obtained according to 
\begin{eqnarray}
T_{\mbox{co-added}} = \frac{\sum_{I=3}^{10} T_I / \sigma^2_{0,I}}
{\sum_{I=3}^{10} 1 / \sigma^2_{0,I}} \, , \nonumber
\end{eqnarray}
where $T_I$ is the CMB map for the DA $I$, and $\sigma^2_{0,I}$ is 
the noise {\it per} observation for the DA $I$.

To investigate whether the MPASH function is independent of the number 
of sub-catalogs $K$ in which the original catalog is divided due to a 
large amount of pixels, we study first the angular correlations in the 
North-Galactic and South-Galactic spherical cap regions of $70^{\circ}$ 
of aberture of the co-added WMAP map.
Our neat results shown in Figure 3, for the analysis of seven 
dif\/ferent partitions of the above mentioned regions of the co-added 
CMB map, demonstrates that the MPASH-minus-EPASH function (and of course 
also the MPASH) is actually independent of $K$. 

Finally, in Figure 4 we show the same result, but this time for the 
full-sky WMAP CMB map, termed the `cleaned' map~\cite{Tegmark}, 
which was produced by combining all the individual frequency WMAP maps 
using a Wiener filter algorithm in order to reduce or eliminate 
foregounds (specially those coming from our Galaxy). 

\begin{figure} 
\includegraphics[width=9cm, height=6.5cm]{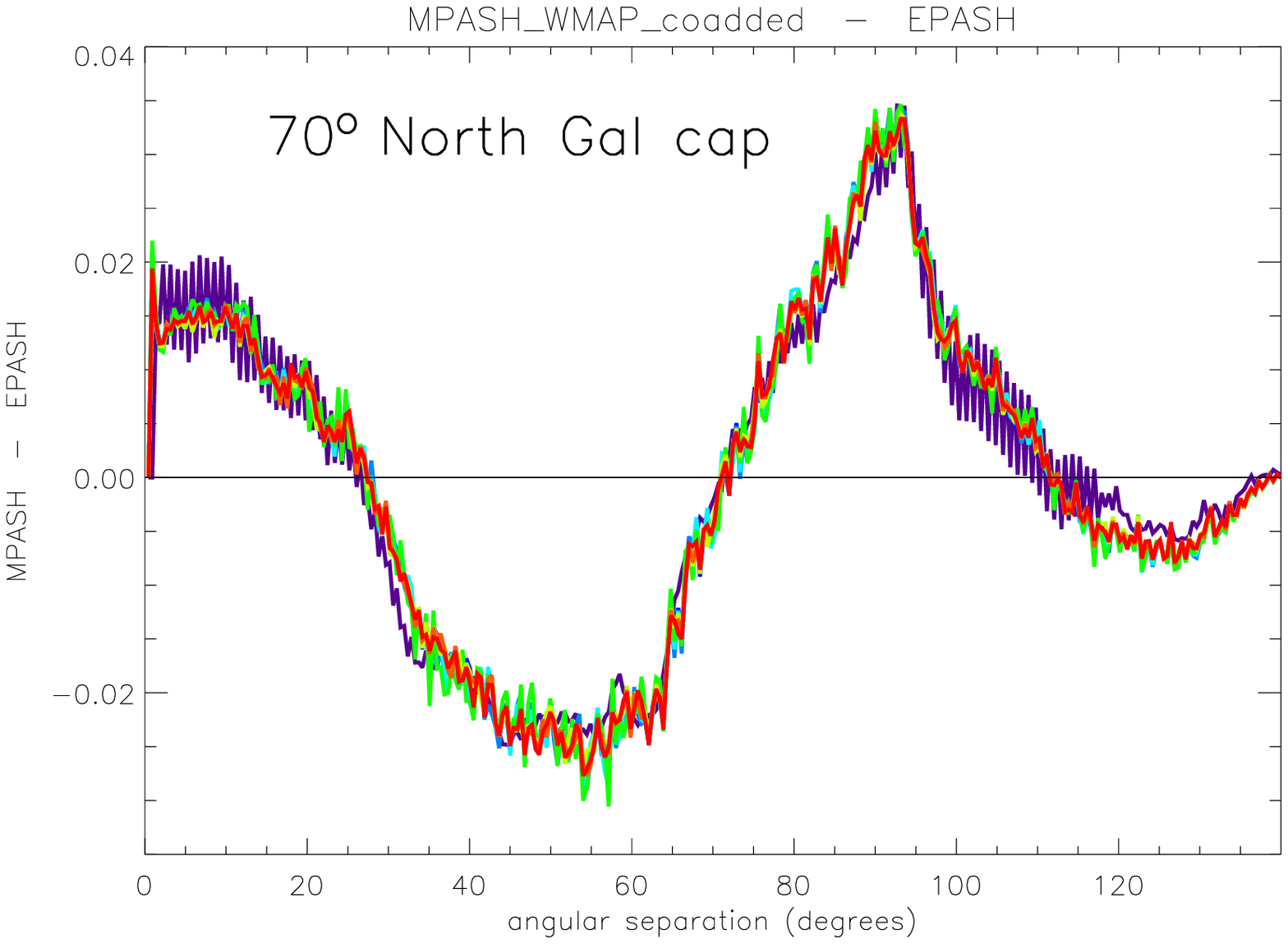}
\includegraphics[width=9cm, height=6.5cm]{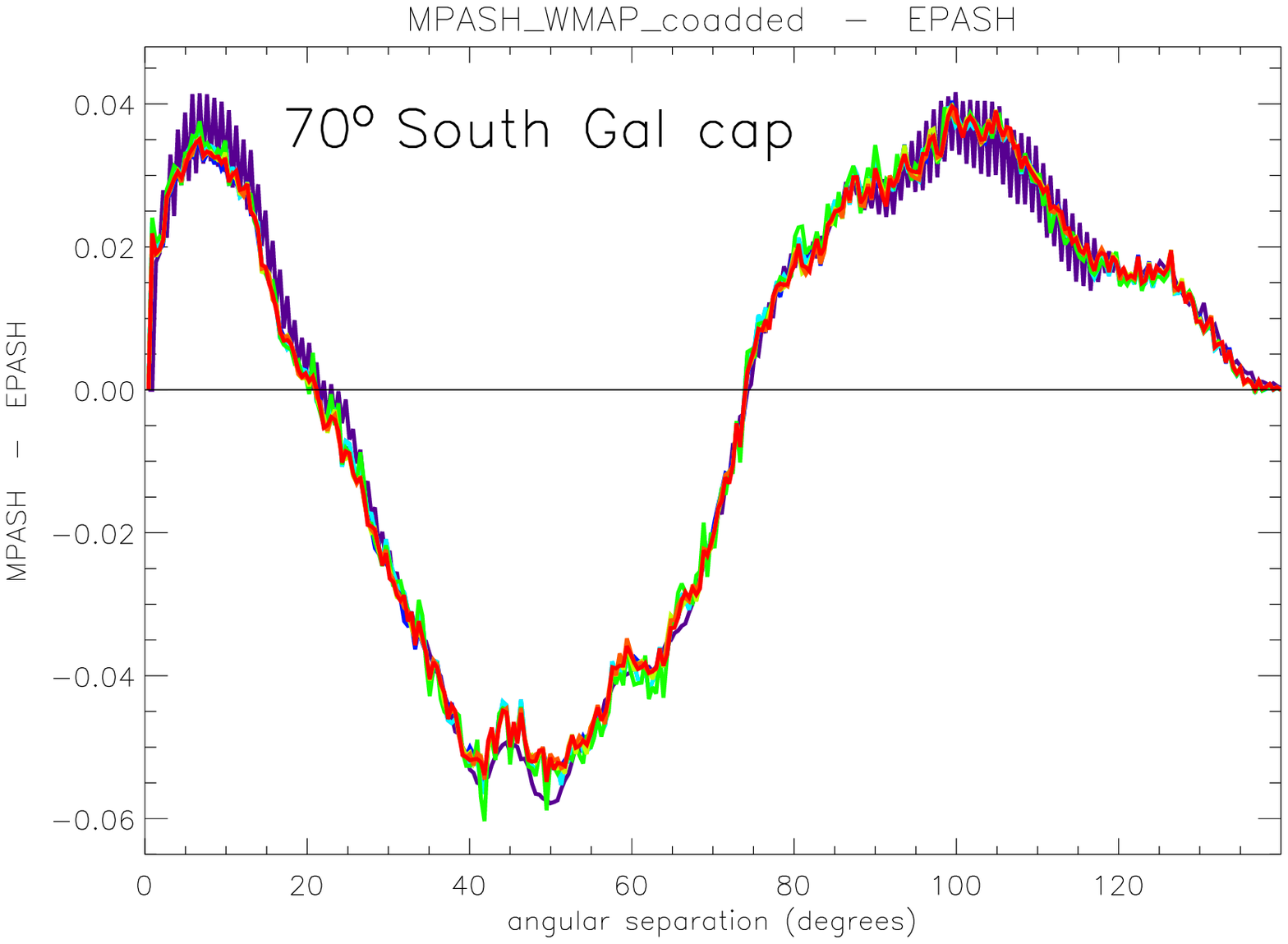}
\caption{ \label{figure3}
MPASH-minus-EPASH functions for CMB data in the $70^{\circ}$ North 
and South Galactic spherical caps of the co-added WMAP map~\cite{WMAP2}. 
The seven curves plotted in each figure (top and bottom) correspond 
to the cases where the number of histograms was: $K=4, 16, 17, 20, 25, 40,$ 
and $80$, respectively. 
The $K=4$ case corresponds to the degraded map $N_{side}=64$, and all 
the others to $N_{side}=128$~\cite{Gorski}. 
As observed, the coincidence of this rainbow of 
curves proves that the MPASH-minus-EPASH is independent of the number 
$K$.}
\end{figure}  

\begin{figure} 
\includegraphics[width=9cm, height=6.5cm]{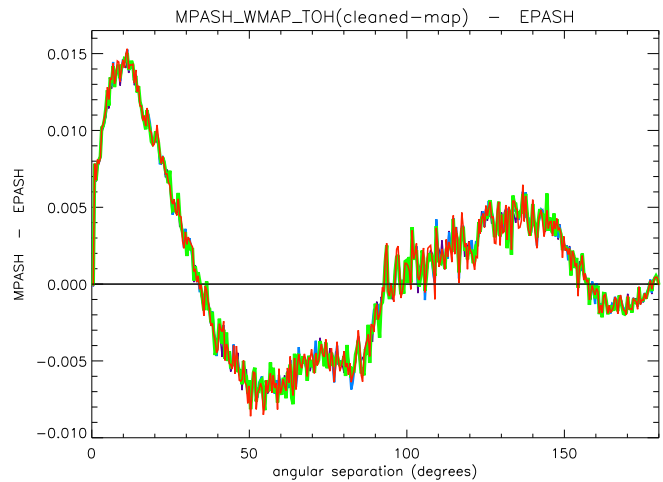}
\caption{ \label{figure4}
MPASH-minus-EPASH functions for the full-sky WMAP `cleaned' 
map~\cite{Tegmark}. 
The four curves plotted here correspond to the cases where the number 
of histograms to construct the MPASH was: $K=40, 81, 100,$ and $130$, 
respectively. 
As observed, all the coloured-curves coincide showing again that
the MPASH-minus-EPASH function is independent of the number $K$.}
\end{figure} 

%
\section{Simple applications of the PASH method}
 
In order to illustrate the MPASH-minus-EPASH function revealing 
signatures in simple cases, here we consider three CMB maps where 
the angular correlations are known {\it a priori}: 
\begin{itemize}
\item 
a statistically isotropic CMB map, that is there are no angular 
correlations between the temperature f\/luctuations of the map; 
the results of our analysis are shown in Figure 5.
\item 
a pure dipole map; the results of our analysis are shown 
in Figure 6.
\item 
a pure quadrupole map; the results of our analysis 
are shown in Figure 7.
\end{itemize}
\begin{figure} 
\includegraphics[width=5cm, height=8cm, angle=90]{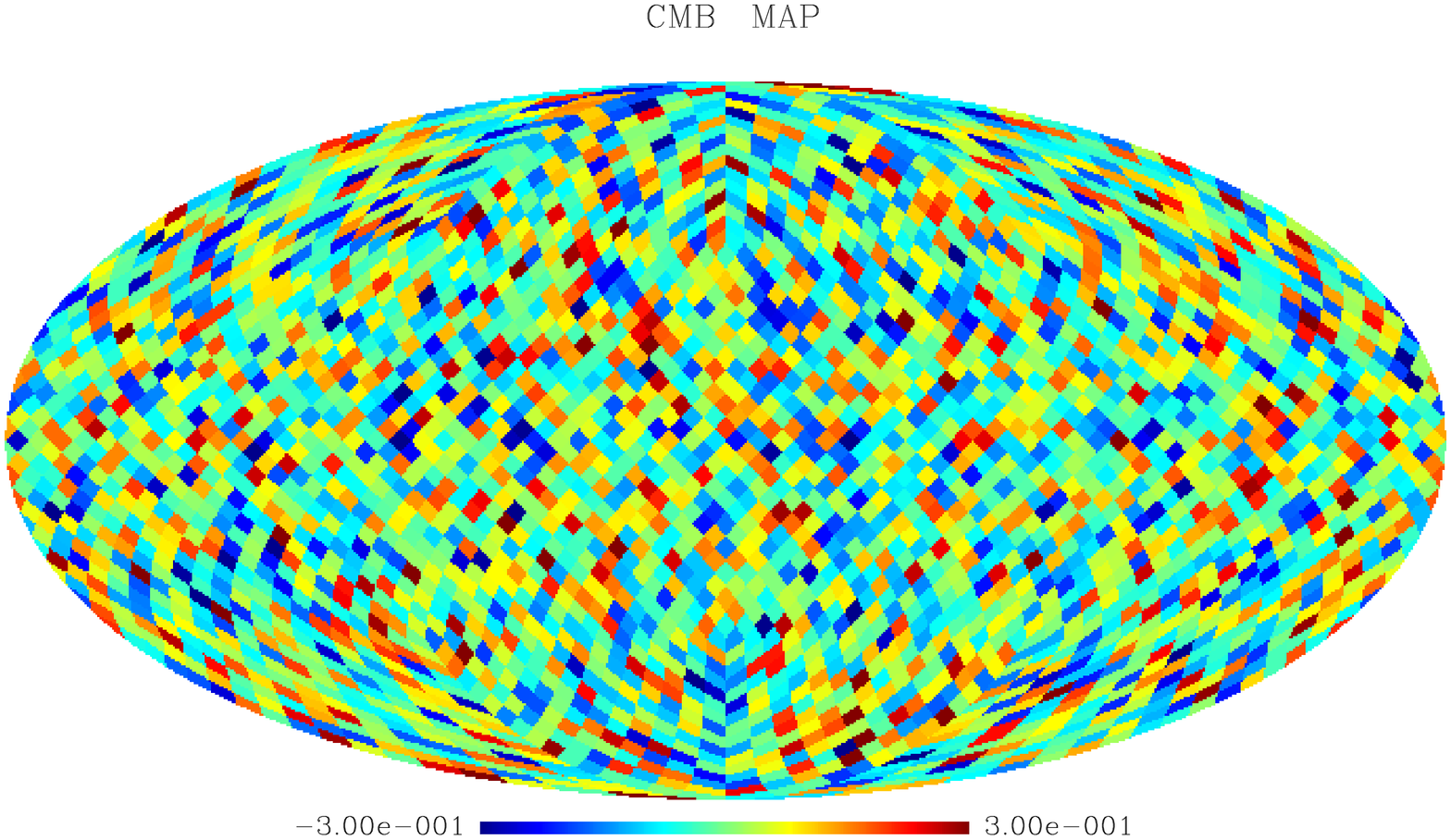}
\includegraphics[width=9cm, height=6cm]{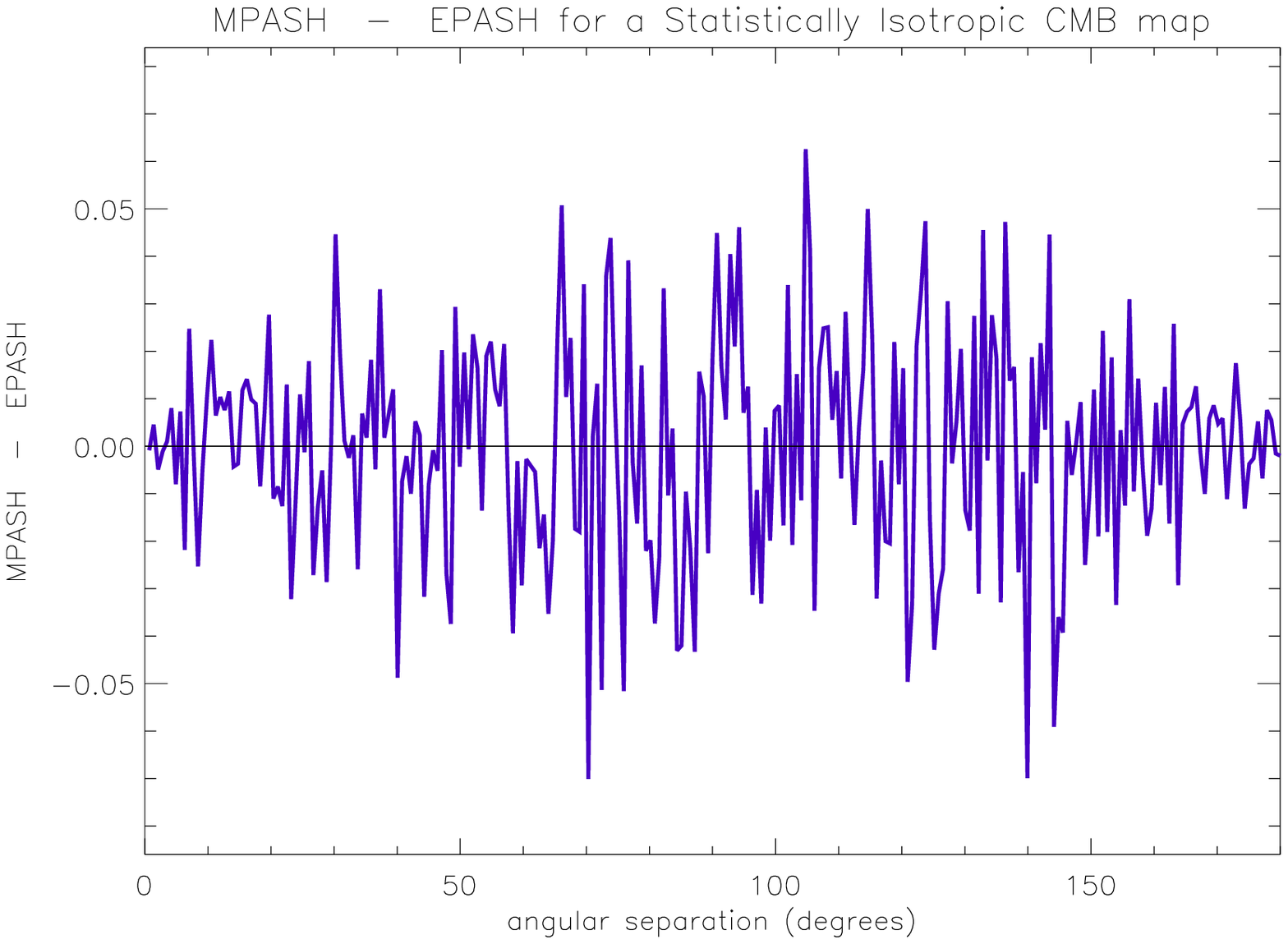}
\caption{ \label{figure5}
Statistically isotropic CMB map and the MPASH-minus-EPASH analysis
for these data. 
The MPASH-minus-EPASH function is just a statistical (noisy) 
oscillations around the horizontal axis, revealing just the absence of 
angular coorelations in the data.} 
\end{figure}   

\begin{figure} 
\includegraphics[width=5cm, height=8.cm, angle=90]{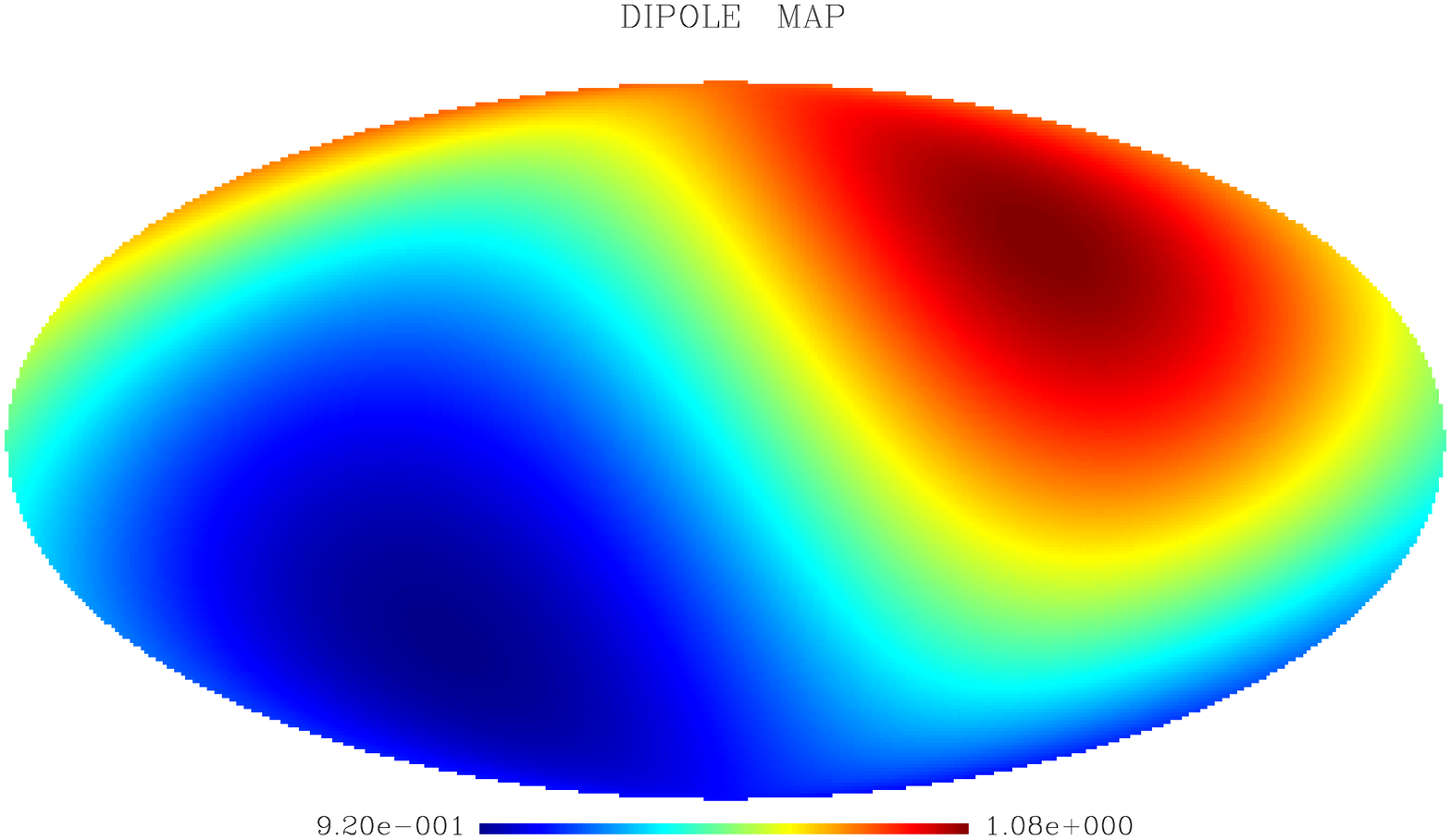}
\includegraphics[width=9cm, height=6.cm]{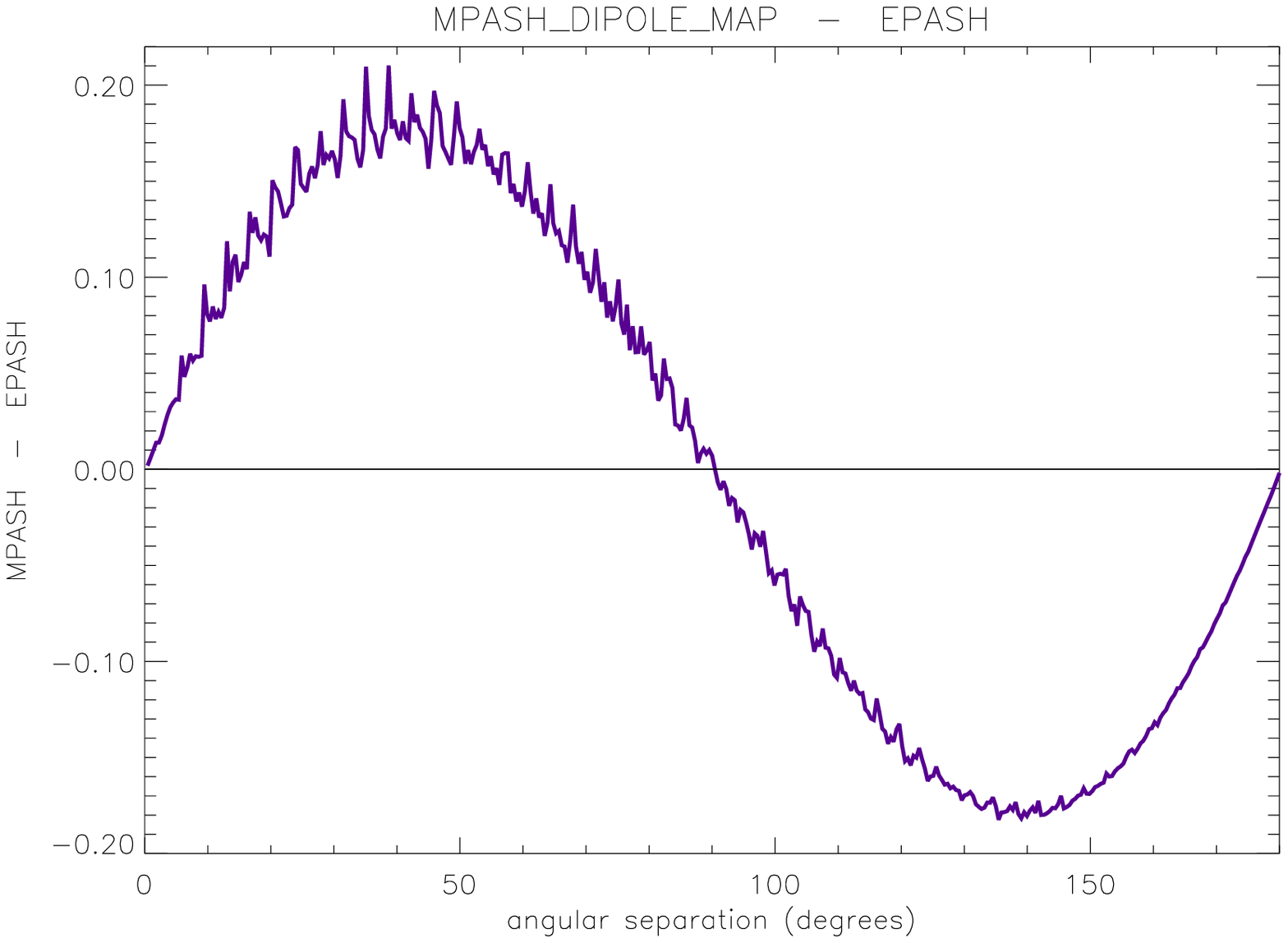}
\caption{ \label{figure6}
A CMB Dipole map and the MPASH-minus-EPASH analysis for these data. 
The MPASH-minus-EPASH function results in a {\em Sine} function, trivially 
revealing a tipical dipole-signature (with some statistical noise).} 
\end{figure}  

\begin{figure} 
\includegraphics[width=5cm, height=8.cm, angle=90]{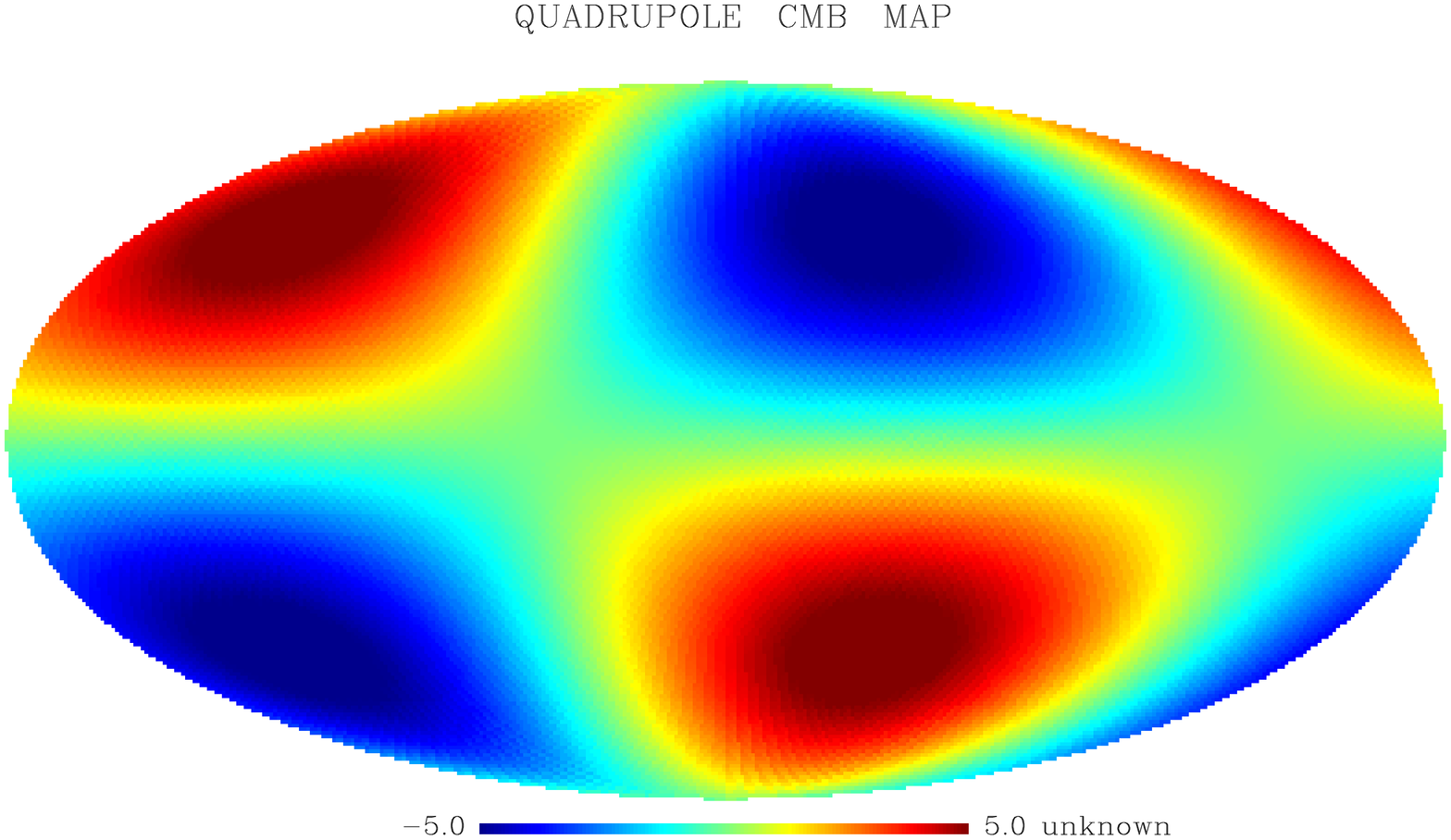}
\includegraphics[width=9cm, height=6.cm]{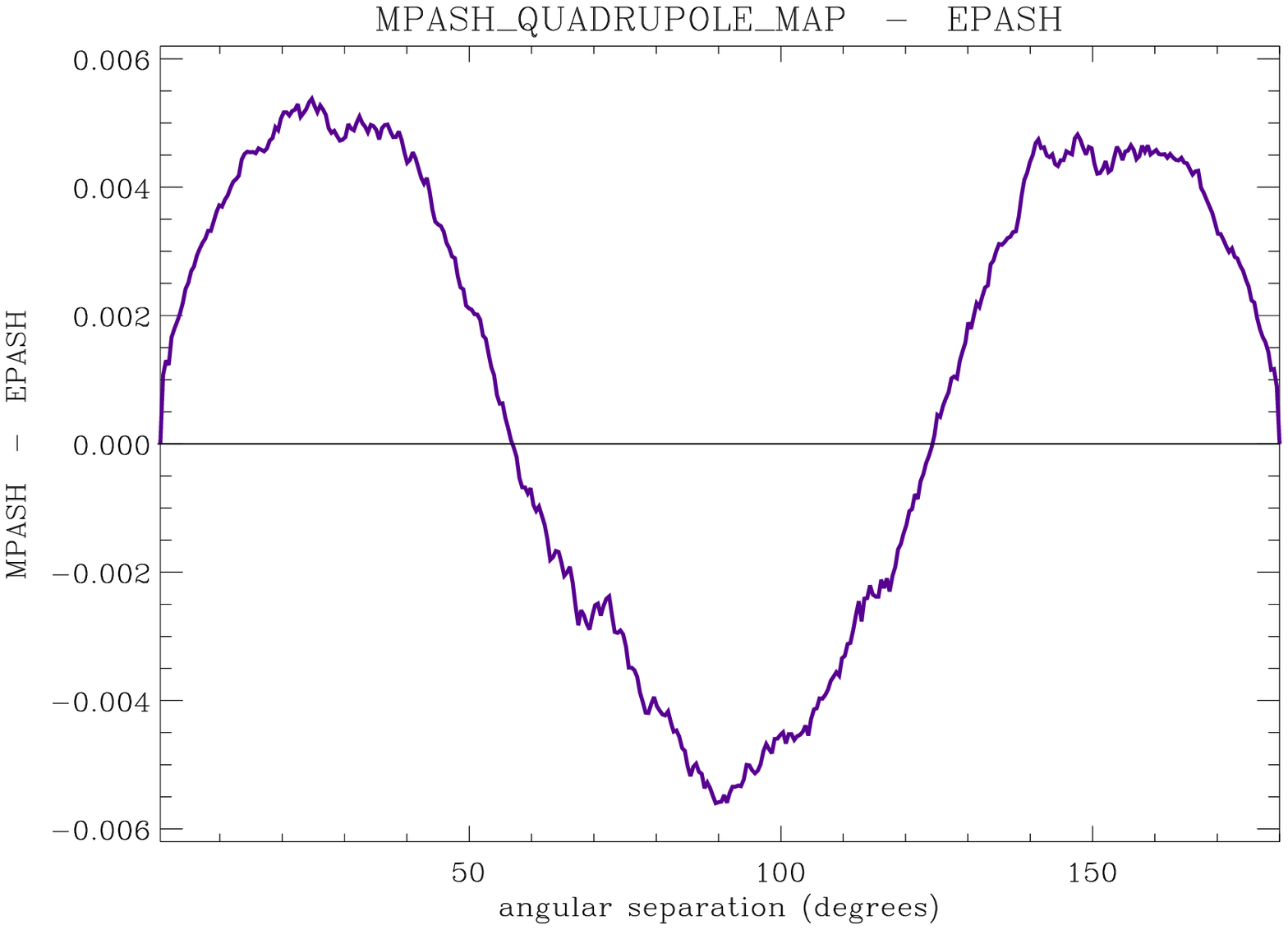}
\caption{ \label{figure7}
A quadrupole CMB map and the MPASH-minus-EPASH function resulting from 
these data.
A short analysis of the physical data let us to realize that this function 
actually represents a quadrupole angular-correlation signature, again with 
some level of statistical noise.} 
\end{figure} 

\section*{Acknowledgements} 
\noindent
I acknowledge use of the Legacy Archive for Microwave Background 
Data Analysis (LAMBDA). 
Some of the results in this paper have been derived using the 
HEALPix~\cite{Gorski} package.
I am grateful to Thyrso Villela for many useful discussions. 
I am indebted to Carlos Alexandre Wuensche, Rodrigo Leonardi, 
and Ivan Ferreira for introducing me into the {\it labyrinthus} 
of the CMB {\sc tools} and the IDL {\it mysterious}. 
I acknowledge also a PCI/7B-CNPq fellowship. 
Last but not least, my thanks to the Organizers of the 
Conference {\sc 100 years of relativity} for their kind hospitality.

\vspace{0.5cm}
\noindent
$^{\ast}$\,bernui\,@\,das.inpe.br


\section{References}

\vspace{-1.75cm}
\begin{thebibliography}{16}
\bibitem[Scott & Smoot(1992)]{COBE} D. Scott and G. F. Smoot, 
astro-ph/0406567 (2004).\\
\bibitem[Hu \& Dodelson(2002)]{Hu} W. Hu and S. Dodelson,
Annual Review of Astronomy and Astrophysics, {\bf 40}, 171 (2002).\\
\bibitem[Hu \& White(2004)]{Hu2} W. Hu and M. White, 
Sci. Amer., Feb, 44 (2004).\\
\bibitem[Bennett et al.(2003)]{WMAP} C. L. Bennett, {\it et al.} 
\apj S. {\bf 148}, 1 (2003).\\
\bibitem[Bernui \& Villela(2005)]{BV} A. Bernui and T. Villela, 
astro-ph/0511339, to appear in A\&A (2005).\\
\bibitem[Bernui, Villela \& Ferreira(2004)]{BVF} A. Bernui, 
T. Villela, and I. Ferreira, Int. J. Mod. Phys. D, {\bf 13}, 
1189 (2004).\\
\bibitem[Bersanelli(2002)]{Bersanelli} M. Bersanelli, D. Maino, 
and A. Mennella, `Anisotropies of the cosmic microwave background', 
Rivista del Nuovo Cimento, Vol.25, N. 9 (2002).\\
\bibitem[Padmanabhan(1993)]{Padmanabhan} T. Padmanabhan, 
`Structure formation in the universe', Cambridge Univ. press, (1993).\\
\bibitem[Hinshaw et al.(2003)]{WMAP2} G. Hinshaw, {\it et al.},
\apj S. {\bf 148}, 135 (2003).\\
\bibitem[Tegmark et al.(2003)]{Tegmark} M. Tegmark, 
A. de Oliveira-Costa, and A. J. S. Hamilton, \prd, {\bf 68}, 
123523 (2003).\\
\bibitem[G\'orski et al.(2005)]{Gorski} 
K. M. G\'orski, {\it et al.}, \apj, 622, 759 (2005). 
\end{thebibliography}
\end{document}